\documentclass[12pt]{article}
\pdfoutput=1
\usepackage{makeidx}
\makeindex
\usepackage[a4paper]{geometry}
\usepackage{jheppub,amsmath,amssymb,amsfonts,amsxtra,mathrsfs,graphics,graphicx,amsthm,epsfig,ytableau,bm,longtable,float,color,tikz,mathtools,xfrac,footnote,rotating,lscape}
\usepackage{dsfont}
\usepackage{textgreek}
\usetikzlibrary{decorations.pathmorphing}
\usetikzlibrary{decorations.markings}
\usetikzlibrary{arrows, decorations.markings, calc, fadings, decorations.pathreplacing, patterns, decorations.pathmorphing, positioning}
\usepackage{tikz-cd}

\usepackage{fixmath} 
\usepackage{scalerel}
\newlength\bshft
\def\fakebold#1{\ThisStyle{\ooalign{$\SavedStyle#1$\cr%
  \kern-\bshft$\SavedStyle#1$\cr%
  \kern\bshft$\SavedStyle#1$}}}

\usetikzlibrary{positioning,shapes}
\usetikzlibrary{chains}
\usetikzlibrary{arrows,fit,decorations.pathreplacing}
\tikzstyle{every picture}+=[remember picture]
\tikzstyle{na} = [baseline=-.5ex]

\usepackage{empheq}
\usepackage{multirow}
\usepackage{booktabs}
\usepackage[american]{babel}

\usepackage[latin1]{inputenc}

\makeatletter
\newcommand{\vast}{\bBigg@{1}}
\newcommand{\Vast}{\bBigg@{5}}
\makeatother

\usepackage{hyperref}

\numberwithin{equation}{section}

\newcommand{\cf}{\textit{cf.}}
\newcommand{\eg}{\textit{e.g.}}

\newcommand{\ie}{\textit{i.e.}}

\numberwithin{equation}{section}

\newcommand{\be}{\begin{equation}} \newcommand{\ee}{\end{equation}}
\newcommand{\bea}{\begin{equation} \begin{aligned}} \newcommand{\eea}{\end{aligned} \end{equation}}

\def\U{\mathrm{U}}
\def\SO{\mathrm{SO}}
\def\O{\mathrm{O}}
\def\SU{\mathrm{SU}}

\def\USp{\mathrm{USp}}

\newcommand{\rd}{\mathrm{d}}

\newcommand{\vol}{\mathrm{vol}}

\newcommand{\wt}{\widetilde}

\newcommand{\cF}{\mathcal{F}}

\newcommand{\cI}{\mathcal{I}}

\newcommand{\cL}{\mathcal{L}}
\newcommand{\cM}{\mathcal{M}}
\newcommand{\cN}{\mathcal{N}}

\newcommand{\cW}{\mathcal{W}}

\newcommand{\fg}{\mathfrak{g}}

\newcommand{\fn}{\mathfrak{n}}

\newcommand{\fs}{\mathfrak{s}}
\newcommand{\ft}{\mathfrak{t}}

\usepackage[Symbol]{upgreek}
\usepackage{bm}

\usepackage{calligra}
\DeclareMathAlphabet{\mathcalligra}{T1}{calligra}{m}{n}

\theoremstyle{plain}

  \theoremstyle{definition}

\providecommand{\examplename}{Example}
\providecommand{\theoremname}{Theorem}

\makeatletter
\g@addto@macro\bfseries{\boldmath}
\makeatother

\title{6D attractors and black hole microstates}

\author[a]{Seyed Morteza Hosseini,}
\author[b]{Kiril Hristov,}
\author[c]{Achilleas Passias,}
\author[d,e]{and Alberto Zaffaroni}
\affiliation[a]{Kavli IPMU (WPI), UTIAS, The University of Tokyo, Kashiwa, Chiba 277-8583, Japan}
\affiliation[b]{Institute for Nuclear Research and Nuclear Energy, Bulgarian Academy of Sciences, \\Tsarigradsko Chaussee 72, 1784 Sofia, Bulgaria}
\affiliation[c]{Department of Physics and Astronomy, Uppsala University, \\Box 516, SE-751 20 Uppsala, Sweden}
\affiliation[d]{Dipartimento di Fisica, Universit\`a di Milano-Bicocca, I-20126 Milano, Italy}
\affiliation[e]{INFN, sezione di Milano-Bicocca, I-20126 Milano, Italy}
\emailAdd{morteza.hosseini@ipmu.jp}
\emailAdd{khristov@inrne.bas.bg}
\emailAdd{achilleas.passias@physics.uu.se}
\emailAdd{alberto.zaffaroni@mib.infn.it}

\preprint{IPMU18-0155, UUITP-41/18}

\abstract{We find a family of AdS$_2\times \cM_4$ supersymmetric solutions of the six-dimensional $\mathrm{F}(4)$ gauged supergravity coupled to one vector multiplet that arises as a low energy description of massive type IIA supergravity on (warped) AdS$_6\times S^4$. $\cM_4$ is either a K\"ahler-Einstein manifold or a product of two Riemann surfaces with a constant curvature metric. These solutions correspond to the near-horizon region of a family of static magnetically charged black holes. In the case where $\mathcal{M}_4$ is a product of Riemann surfaces, we successfully compare their entropy to a microscopic counting based on the recently computed topologically twisted index of the five-dimensional $\mathcal{N}=1$ $\mathrm{USp}(2 N)$ theory with $N_f$ fundamental flavors and an antisymmetric matter field. Furthermore, our results suggest that the near-horizon regions exhibit an attractor mechanism for the scalars in the matter coupled $\mathrm{F}(4)$ gauged supergravity, and we give a proposal for it.}

\begin{document}

\setcounter{tocdepth}{2}
\maketitle

%
%

\date{Dated: \today}




\section{Introduction}
\label{sec:intro}

Recently there has been some progress in understanding the microscopic origin of the Bekenstein-Hawking entropy of supersymmetric asymptotically anti-de Sitter (AdS) black holes.
In particular, the microscopic entropy of certain four-dimensional static, dyonic, BPS black holes \cite{Cacciatori:2009iz,DallAgata:2010ejj,Hristov:2010ri,Katmadas:2014faa,Halmagyi:2014qza},
which can be embedded in AdS$_4 \times S^7$, has been reproduced by a field theory calculation based on the topologically twisted index \cite{Benini:2015noa}
of the dual ABJM theory in the large $N$ limit \cite{Benini:2015eyy,Benini:2016rke,Benini:2016hjo}.
These black holes have an AdS$_2 \times \Sigma_{\fg}$ near-horizon geometry, where $\Sigma_{\fg}$ is a Riemann surface of genus $\fg$.
The topologically twisted index is the partition function of the dual field theory on $\Sigma_{\fg} \times S^1$, partially topologically $A$-twisted along $\Sigma_{\fg}$.
Specifically, the index $Z (p_I, \Delta_I)$ is a function of a set of magnetic charges $p_I$ and complexified chemical potentials $\Delta_I$ for the global symmetries of the theory.
The statistical entropy $S_{\text{BH}}$ of the black holes with purely magnetic charges is then obtained by evaluating $Z (p_I,\Delta_I)$ at its critical point $\bar \Delta_I$:
\be
 \cI_{\text{SCFT}}  ( p_I , \bar \Delta_I ) \equiv \log Z (p_I, \bar \Delta_I)  = S_{\text{BH}} (p_I) \, .
\ee
This procedure was dubbed $\cI$-extremization in \cite{Benini:2015eyy}. These results have been generalized to different black holes in four dimensions and black strings in five dimensions
\cite{Hosseini:2016tor,Hosseini:2016ume, Hosseini:2016cyf,Hosseini:2018qsx,Hong:2018viz,Cabo-Bizet:2017jsl,Azzurli:2017kxo,Hosseini:2017fjo,Benini:2017oxt,Bobev:2018uxk,Toldo:2017qsh,Gang:2018hjd}.
For other interesting progresses in this context see \cite{Liu:2017vbl,Liu:2017vll,Jeon:2017aif, Halmagyi:2017hmw,Cabo-Bizet:2017xdr, Hristov:2018lod, Nian:2017hac,Hosseini:2017mds,Hosseini:2018dob,Liu:2018bac}.

The five-dimensional topologically twisted index, which is the partition function of a five-dimensional ${\cal N}=1$ theory on $\cM_4\times S^1$ with an Abelian topological twist along $\cM_4$, has been recently computed
when $\cM_4$ is a toric K\"ahler manifold \cite{Hosseini:2018uzp} or the product of two Riemann surfaces \cite{Hosseini:2018uzp,Crichigno:2018adf}.
Therein, the ${\cal N}=1$ $\USp(2 N)$ gauge theory with $N_f$ hypermultiplets in the fundamental representation
and one hypermultiplet in the antisymmetric representation of $\USp(2N)$, arising on the worldvolume of D4-branes near D8-branes and orientifolds \cite{Seiberg:1996bd}, has been analyzed in the large $N$ limit.
With some assumptions on the relevant saddle-point, the large $N$ limit of the index for $\cM_4 = \Sigma_{\fg_1}\times \Sigma_{\fg_2}$ has been evaluated as a function of magnetic charges and chemical potentials
for the Cartan subgroup of the $\SU(2)_M$ global symmetry of the theory \cite{Hosseini:2018uzp}.%
\footnote{A different proposal was discussed in  \cite{Crichigno:2018adf}. In this paper we use the proposal in \cite{Hosseini:2018uzp} that nicely matches the entropy of the black holes.}
This result provides a prediction for the entropy of a family of AdS$_6$ magnetically charged black holes in massive type IIA supergravity.
This prediction has been successfully tested for the only existing black hole solution, the so-called universal one \cite{Benini:2015bwz,Bobev:2017uzs} with a particular value of the magnetic charge,
using the results for the entropy given in \cite{Suh:2018tul}. It is the purpose of this paper to find new black hole solutions, explicitly depending on a set of magnetic charges,
and show that their entropy is correctly accounted by the topologically twisted index.    

To this end, we will consider a six-dimensional truncation of the supersymmetric warped AdS$_6 \times S^4$  background of massive type IIA supergravity \cite{Brandhuber:1999np} dual to the $\USp(2 N)$ theory.
This truncation is described by an ${\rm F}(4)$ gauged supergravity coupled to vector multiplets \cite{Ferrara:1998gv,Romans:1985tw,DAuria:2000xty,Andrianopoli:2001rs}.
Furthermore, we will restrict ourselves to one vector multiplet corresponding to the Cartan subgroup of the $\SU(2)_M$ global symmetry of the five-dimensional superconformal field theory (SCFT).

As a warm-up, and to test the consistency of the truncation, we consider the background AdS$_4\times \Sigma_{\fg}$ which corresponds to a twisted compactification of the five-dimensional field theory on $\Sigma_{\fg}$.
We successfully compare the free energy of the solution with the field theory computation in \cite{Crichigno:2018adf} and the ten-dimensional gravity computation in \cite{Bah:2018lyv}.

We then find new black hole horizon geometries of the form AdS$_2 \times \Sigma_{\fg_1} \times \Sigma_{\fg_2}$. 
We turn on an Abelian gauge field inside the $\SU(2)$ R-symmetry that performs the topological twist by cancelling the spin connection, and two magnetic fluxes $p_1$ and $p_2$ (one along each Riemann surface) for the $\U(1)$ gauge field in the additional vector multiplet. We thus have a two-parameter family of magnetically charged black holes. We compare the entropy with the value of the topologically twisted index $Z(p_1,p_2,\Delta)$ that also depends on a chemical potential for the $\U(1) \subset \SU(2)_M$ symmetry and we find that the statistical entropy $S_{\text{BH}}$ of the black holes as a function of the magnetic charges
is obtained by evaluating $Z (p_1,p_2,\Delta)$ at its critical point $\bar \Delta$:
\be
 \label{intro:I-extremization}
 \cI_{\text{SCFT}}  ( p_1, p_2 , \bar \Delta ) \equiv \log Z (p_1, p_2 , \bar \Delta)  = S_{\text{BH}} (p_1 , p_2) \, .
\ee
With a convenient \emph{democratic} parameterization%
\footnote{The $\Delta_I$, $I = 1, 2$, parameterize the Cartan of the $\SU(2)_R$ and the $\SU(2)_M$ symmetry of the $\USp(2N)$ theory.
They satisfy the constraint \eqref{dem}. Similarly, one can introduce a redundant, but democratic, parameterization for the fluxes as in \eqref{parfluxes}.
With such a choice, the topologically twisted index is a homogeneous function of $\Delta_I$, $\fs_I$ and $\ft_I$. See \cite{Hosseini:2018uzp} for details.}
for the fluxes and chemical potentials the explicit form of the index can be written as \cite{Hosseini:2018uzp}
\be\label{summary}
 \cI_{\text{SCFT}} (\fs_I, \ft_I, \Delta_I) = \frac{4 \sqrt{2}}{15} \frac{N^{5/2}}{\sqrt{8 - N_{f}}} \sum_{I , J = 1}^2 \fs_I \ft_J \frac{\partial^2 ( \Delta_1 \Delta_2)^{3/2}}{\partial \Delta_I \partial \Delta_J} \, .
\ee
This structure is reminiscent of an analogous result for AdS$_4$ black holes \cite{Benini:2015eyy,Benini:2016rke,Benini:2016hjo,Hosseini:2016tor,Hosseini:2017fjo,Benini:2017oxt}.
This analogy and the relation to other interesting field theory quantities like the $S^5$ free energy and the effective twisted superpotential
of the partial compactification on one of the Riemann surfaces, were discussed in detail in \cite{Hosseini:2018uzp}. 

We also obtain AdS$_2 \times \cM_4$ horizon geometries where $\cM_4$ is a four-dimensional K\"ahler-Einstein manifold depending on a magnetic flux along $\cM_4$.
We find a simple and intriguing expression for the entropy suggesting that the computation in \cite{Hosseini:2018uzp} could be generalized to this case too. We leave this for future work.

In gravity, the field theory chemical potential $\Delta$ can be associated with the horizon value of the  vector multiplet scalar field $\phi_3$.
With a convenient parameterization, we find that the functional $\cI_{\text{SCFT}} ( p_1, p_2 , \bar \Delta )$ coincides with
the area of the horizon divided by $4 G_{\text{N}}$, where $G_{\text{N}}$  is the six-dimensional Newton's constant, as a function of 
$\phi_3$. This is the {\it attractor mechanism} in six-dimensional gauged supergravity: after expressing all the fields
in the gravity multiplet in terms of vector multiplet scalars using the BPS equations,
the remaining BPS equations are equivalent to the extremization of the area of the horizon as a functional of vector multiplet scalars,
and the critical value of this functional is the entropy.
We see that the $\cI$-extremization principle is equivalent to the attractor mechanism in six-dimensional gauged supergravity,%
\footnote{For the attractor mechanism in six-dimensional {\it ungauged} supergravity, instead, see \eg\;\cite{Ferrara:2008xz,Lam:2018jln} and references therein.}
thus generalizing what was found for AdS$_{4}$ black holes in \cite{Benini:2015eyy,Benini:2016rke,Benini:2016hjo,Hosseini:2017fjo,Benini:2017oxt}.
 
More explicitly, we find that a central role is played by the quantity \eqref{prepotential}
\be\label{prepotential0}
 \cI_{\text{AdS}_6} (X^I)  = -  \frac{1}{3 \pi G_{\text{N}}} (X^1 X^2 )^{3/2} \, ,
\ee
where $X^I(\phi_3)$ $(I = 1 , 2)$, defined in \eqref{parametrization}, are the gravity counterpart of $\Delta_I$ in \eqref{summary}.
This six-dimensional quantity is reminiscent and can be thought of as the analogue of the prepotential ${\cal F}_{\text{sugra}} (X^I)$ in four-dimensional $\cN = 2$ gauged supergravity.
Indeed, we will find that the attractor equations for AdS$_4$ vacua correspond to extremizing
\be
 \label{attr1}
 \cI_{\text{AdS}_4} (X^I) = \frac{8 \pi}{27} \sum_{I = 1}^2 \fs_I \frac{\partial \cI_{\text{AdS}_6} (X^I)}{\partial X^I} \, ,
\ee
and attractor equations for black holes correspond to extremizing
\be
 \label{attr2}
 \cI_{\text{AdS}_2} (X^I) = - \frac{\vol (\cM_4)}{108} \sum_{I , J =1}^2 \fs_I \fs_J \frac{\partial^2 \cI_{\text{AdS}_6} (X^I)}{\partial X^I \partial X^J} \, ,
\ee
for $\cM_4$ being a K\"ahler-Einstein manifold, and to extremizing
\be
\label{attr3}
 \cI_{\text{AdS}_2} (X^I) = - \frac{4 \pi^2}{27} \sum_{I , J =1}^2 \fs_I \ft_J \frac{\partial^2 \cI_{\text{AdS}_6} (X^I)}{\partial X^I \partial X^J} \, ,
\ee
for $\cM_4 = \Sigma_{\fg_1} \times \Sigma_{\fg_2}$.
Here $\fs_I$ and $\ft_I$ are the magnetic charges --- see \eqref{fluxesAdS4}, \eqref{parfluxes} and \eqref{fluxesM4}.
This is similar to the attractor mechanism in four-dimensional $\cN = 2$ gauged supergravity \cite{Cacciatori:2009iz,DallAgata:2010ejj}.
We thus expect that, in more general ${\rm F}(4)$ gauged supergravites coupled to vector multiplets, the attractor equations
for AdS solutions supported by magnetic fluxes are given by extremizing expressions of the form \eqref{prepotential0}-\eqref{attr3}
with a suitable function $\cI_{\text{AdS}_6} (X^I)$, homogeneous of degree three.

The structure of this paper is as follows. In section \ref{sec:F(4):sugra} we discuss general aspects of the ${\rm F}(4)$ gauged supergravity coupled to vector multiplets.
In section \ref{AdS6} we discuss the AdS$_6$ vacuum and an interesting partially off-shell version of its free energy that we relate to its field theory counterpart.
In section \ref{AdS4} we consider the background AdS$_4\times \Sigma_{\fg}$ with a topological twist on $\Sigma_{\fg}$ and we successfully compare the free energy
of the solution with the field theory computation in \cite{Crichigno:2018adf} and the ten-dimensional gravity computation in \cite{Bah:2018lyv}.
In section \ref{AdS2} we obtain a two-parameter family of black hole horizons AdS$_2 \times \Sigma_{\fg_1} \times \Sigma_{\fg_2}$ and successfully reproduce their entropy using the topologically twisted index.
In section \ref{AdS2M} we find a one-parameter family of black hole horizons AdS$_2 \times \cM_4$ where $\cM_4$ is a four-dimensional K\"ahler-Einstein manifold.
Our conventions and some useful formulae are collected in appendix \ref{sec:app}.

\paragraph*{Note added:} While we were writing this work, we became aware of  \cite{Suh:2018szn} which  has some overlaps with the results presented here.

\section[Matter coupled \texorpdfstring{${\rm F}(4)$}{F(4)} gauged supergravity]{Matter coupled ${\rm F}(4)$ gauged supergravity}
\label{sec:F(4):sugra}

We consider a six-dimensional truncation of the supersymmetric warped AdS$_6 \times S^4$  background of massive type IIA supergravity \cite{Brandhuber:1999np} described by an ${\rm F}(4)$ gauged supergravity coupled to vector multiplets.
The minimal ${\rm F}(4)$ gauged supergravity was written in \cite{Romans:1985tw} and coupled to matter in \cite{DAuria:2000xty,Andrianopoli:2001rs}.
${\rm F}(4)$ is the relevant superalgebra for five-dimensional superconformal field theories and its bosonic subalgebra is $S\O(5,2) \times \SU(2)_R$.

The bosonic part of the six-dimensional gravity multiplet consists of the metric $g_{\mu\nu}$,
four vectors $A^\alpha$, $\alpha=0,1,2,3$, a two-form $B_{\mu\nu}$ and the dilaton $\sigma$.
It is useful to split $\alpha=(0,r)$ where $r=1,2,3$ is an index in the adjoint representation of $\SU(2)_R$.
The fermionic components are a gravitino $\psi_\mu^A$ and a spin one-half fermion $\chi^A$, $A=1,2$,  transforming in the fundamental representation of $\SU(2)_R$.

The vector multiplet in six-dimensions contains a gauge field $A_\mu$, four scalars $\phi_\alpha$ and  a spin one-half fermion $\lambda_A$.
With $n_\text{V}$ vector multiplets, the $4 n_\text{V}$ scalar fields parameterize the coset space
\be
 \frac{\SO(4,n_\text{V})}{\SO(4)\times \SO(n_\text{V})} \, .
\ee
It is convenient to encode the scalar fields into a coset representative  $L^\Lambda_{\phantom{\Lambda}\Sigma} \in \SO(4,n_\text{V})$, where indices are split as follows $\Lambda=(\alpha,I)$
with $I=1,\ldots n_\text{V}$. A subgroup $\SU(2)_R \times G$ of dimension $3 + n_\text{V}$ of $\SO(4,n_\text{V})$ can be gauged.

The bosonic Lagrangian reads \cite{Andrianopoli:2001rs}%
\footnote{We follow the conventions of \cite{Andrianopoli:2001rs}. Notice that \cite{Andrianopoli:2001rs} employs the unusual convention $F = F_{\mu\nu} \rd x^\mu \wedge \rd x^\nu$ for the components of a form.
In particular, for them $F_{\mu\nu} = \frac 12 \left( \partial_\mu A_\nu - \partial_\nu A_\mu \right)$.} 
\bea
 \label{Lagrangian}
 \cL = & - \frac14 R -\frac18 e^{-2 \sigma}{\cal N}_{\Lambda\Sigma} \hat F_{\mu\nu}^\Lambda \hat F^{\Sigma\mu\nu} + \frac{3}{64} e^{4 \sigma} H_{\mu\nu\rho}H^{\mu\nu\rho} + \partial^\mu \sigma \partial_\mu \sigma -\frac14 P^{I\alpha\mu} P_{I\alpha\mu} - V \\
 & - \frac{1}{64} \epsilon^{\mu\nu\rho\sigma\lambda\tau} B_{\mu\nu} \left( \eta_{\Lambda\Sigma} \hat F_{\rho\sigma}^\Lambda \hat F_{\lambda\tau}^\Sigma + m B_{\rho\sigma}  \hat F_{\lambda\tau}^0 +\frac13 m^2 B_{\rho\sigma}  B_{\lambda\tau} \right) \, ,
\eea 
where
\bea
 & \hat F_{\rho\sigma}^\Lambda = F_{\rho\sigma}^\Lambda - m \delta^{\Lambda 0} B_{\mu\nu} \, , \\
 & {\cal N}_{\Lambda\Sigma} = L_\Lambda^{\phantom{\Lambda}\alpha} ( L^{-1})_{\alpha\Sigma} - L_\Lambda^{\phantom{\Lambda}I} ( L^{-1})_{I\Sigma} \, , \\
 & P^I_\alpha = (L^{-1})^I_{\phantom{I}\Lambda} \left( \rd L^\Lambda_{\phantom{\Lambda}\alpha} - f_{\Gamma\phantom{\Lambda}\Pi}^{\phantom{\Pi}\Lambda} A^\Gamma L^\Pi_{\phantom{\Pi}\alpha} \right) \, ,
\eea
with $f^\Lambda_{\phantom{\Lambda}\Pi\Gamma}$ the structure constants of the gauge group $\SU(2)_R\times G$.
Here, $g$ is the gauge coupling constant and $m$ is the mass parameter of the massive type IIA supergravity \cite{Cvetic:1999un}.

The supersymmetry variations of the fermions are given by
\bea
 \label{BPS}
 \delta \psi_{A\mu} & = \nabla_\mu \epsilon_A - \frac i2 g \sigma^r_{AB} A_{r\mu} \epsilon^B +\frac{1}{16} e^{-\sigma} \left[ \hat T_{[AB]\nu\lambda} \gamma_7 - T_{(AB)\nu\lambda} \right]
 \left( \gamma_\mu^{\phantom{\mu}\nu\lambda} - 6 \delta_\mu^\nu \gamma^\lambda \right) \epsilon^B \\
 & + \frac{i}{32} e^{2\sigma} H_{\nu\lambda\rho} \gamma_7 \left( \gamma_\mu^{\phantom{\mu} \nu\lambda\rho} - 3 \delta_\mu^\nu \gamma^{\lambda\rho} \right) \epsilon_A + S_{AB} \gamma_\mu \epsilon^B \, , \\
 \delta \chi_A & = \frac{i}{2} \gamma^\mu \partial_\mu \sigma \epsilon_A +\frac{i}{16} e^{-\sigma} \left[ \hat T_{[AB]\nu\lambda} \gamma_7 + T_{(AB)\nu\lambda} \right] \gamma^{\nu\lambda} \epsilon^B
 + \frac{1}{32} e^{2\sigma} H_{\nu\lambda\rho} \gamma_7 \gamma^{\nu\lambda\rho}\epsilon_A + N_{AB} \epsilon^B \, , \\
 \delta \lambda^I_A & = i P^I_{r\mu}  \sigma^{r}_{AB} \gamma^\mu \epsilon^B - i P^I_{0\mu} \epsilon_{AB} \gamma^7\gamma^\mu \epsilon^B + \frac i2 e^{-\sigma} T^I_{\mu\nu} \gamma^{\mu\nu} \epsilon_A +M_{AB}^I \epsilon^B \, ,
\eea
where we suppressed the quadratic terms  in fermions, $\sigma^{rA}_{\phantom{rA} B}$ are the Pauli matrices, and we have defined
\bea \label{T-tensor}
 \hat T_{[AB]\nu\lambda}=\epsilon_{AB} L^{-1}_{0\Lambda} \hat F^\Lambda_{\nu\lambda} \, ,\qquad
 T_{(AB)\nu\lambda}=\sigma^r_{AB} L^{-1}_{r\Lambda} F^\Lambda_{\nu\lambda} \, , \qquad
 T_{I\nu\lambda} = L^{-1}_{I\Lambda} F^\Lambda_{\nu\lambda} \, .
\eea
In all the above formulae the indices $\Lambda, \Pi, \Gamma, \ldots$ are raised and lowered with  the  $\SO(4,n_\text{V})$ invariant metric
$\eta_{\Lambda\Sigma} ={\rm diag}\{ 1,1,1,1,-1,\dots, -1\}$ and the indices $A,B, \ldots$ with the $\SU(2)_R$ tensor $\epsilon_{AB}$.
We refer to the appendix for conventions, for the explicit form of the potential $V$, and the fermion mass matrices $S_{AB},N_{AB},M_{AB}^I$ appearing in \eqref{BPS}.

The five-dimensional superconformal field theory dual to the warped background AdS$_6 \times S^4$ has a gauge group $\USp(2N)$, $N_f$ hypermultiplets in the fundamental representation
and one hypermultiplet in the antisymmetric representation. The theory has an $\SU(2)_R \times \SU(2)_M \times \SO(2 N_f) \times \U(1)_I$ symmetry \cite{Seiberg:1996bd}.%
\footnote{This is non-perturbatively enhanced to $\SU(2)_R \times \SU(2)_M \times {\rm E}_{N_f+1}$.}
The global $\SU(2)_M$ acts on the antisymmetric field, $\SO(2 N_f)$ on the fundamentals and $\U(1)_I$ is the conserved instanton current.

We thus just consider a supergravity containing one vector multiplet, $n_\text{V}=1$, corresponding to the $\U(1)$ subgroup of the global $\SU(2)_M$.
We will consistently set to zero all gauge fields except $A^{r=3}_\mu$ in $\SU(2)_R$ and $A^{I=1}_\mu$ that are needed for the twisting
and to provide magnetic charges for the black holes. We will also require the scalar fields in the vector multiplet $\phi_{\alpha}$, $\alpha = 0 ,1 , 2, 3$,
to be neutral under  $A^{r=3}_\mu$ and this restricts the nonzero components
to $\phi_0$ and $\phi_3$. For purely magnetic black holes we can find solutions with $\phi^0 = 0$ and we further restrict to this case.%
\footnote{One can think of the $\phi_0 = 0$ as analogous to the vanishing of the axions for the AdS$_4$ magnetic black holes \cite{Cacciatori:2009iz,DallAgata:2010ejj,Hristov:2010ri}.}
A convenient parameterization of the scalar coset is given by \cite{Karndumri:2015eta,Gutperle:2017nwo,Gutperle:2018axv}
\be
 L^\Lambda_{\phantom{\Lambda} \Sigma} =
 \begin{pmatrix}
  1 & 0 & 0 & 0 & 0 \\ 0 & 1 & 0 & 0 & 0 \\ 0 & 0 & 1 & 0 & 0 \\ 0 & 0 & 0 & \cosh (\phi_3) & \sinh (\phi_3) \\ 0 & 0 & 0 & \sinh (\phi_3) & \cosh (\phi_3)
 \end{pmatrix} \, .
\ee
The kinetic terms for the vectors can then be written as
\be {\cal N}_{\Lambda\Sigma} =
 \begin{pmatrix}
  1 & 0 & 0 & 0 & 0 \\ 0 & 1 & 0 & 0 & 0 \\ 0 & 0 & 1 & 0 & 0 \\ 0 & 0 & 0 & \cosh (2\phi_3) & - \sinh (2\phi_3) \\ 0 & 0 & 0 & -\sinh (2\phi_3) & \cosh (2\phi_3)
 \end{pmatrix} \, ,
\ee 
and the quantities in the fermionic variations read
\bea
 S_{AB} & = \frac i4 \left( g \cosh (\phi_3) e^\sigma + m e^{-3\sigma} \right) \epsilon_{AB} \, , \\
 N_{AB} & = \frac 14 \left( g \cosh (\phi_3) e^\sigma -3 m e^{-3\sigma} \right) \epsilon_{AB} \, , \\ 
 M_{AB} & = -2 g \sinh (\phi_3) e^\sigma \sigma^3_{AB} \, .
\eea
The other fields that are turned on are the metric, the dilaton $\sigma$ and the two-form $B_{\mu\nu}$.
It is consistent to set $H_{\mu\nu\lambda}=0$ but $B_{\mu\nu}$ is \emph{not} in general zero and its value can be found by solving its equations of motion \cite{Suh:2018tul}.

We believe that after all this simplification the theory is a consistent truncation of massive type IIA supergravity on the warped background AdS$_6 \times S^4$. We give evidence for this in section \ref{AdS4} where we match the ten-dimensional result found in \cite{Bah:2018lyv}. 

We finish the discussion of the matter coupled theory with an argument about the definition of the R-symmetry for all asymptotically AdS$_6$ solutions in the theory. Let us first recall that the detailed match between supergravity and field theory for asymptotically AdS$_4$ black holes was facilitated by the gravitational answer for the R-symmetry
along the holographic renormalization group (RG) flow \cite{Benini:2015eyy}, telling us explicitly how the R-symmetry mixing is parametrized by the values of the scalar fields. In four-dimensional gauged supergravity
the R-symmetry was carefully derived via the Dirac bracket of the supercharges ${\cal Q}$ obtained from the Noether procedure \cite{Hristov:2011ye,Hristov:2011qr}.
Following rigorously all these steps in six dimensions is out of our scope here; however, we can still provide some solid arguments and derive the expected gravitational R-symmetry
as an explicit scalar dependent combination of the two $\U(1)$'s mixing along the flow, $F^{r=3}_{\mu\nu}$ and $F^{I=1}_{\mu\nu} = F^{\Lambda=4}_{\mu\nu}$.
This proposal is strongly backed up by the agreement with the field theory results we provide in the following sections.

It is reasonable to expect that, in analogy to the four-dimensional arguments in \cite{Hristov:2011ye,Hristov:2011qr},
the anti-commutator between two supercharges for asymptotically AdS$_6$ solutions is given by a surface integral%
\footnote{We are evaluating the Dirac bracket of two conserved asymptotic supercharges.
Therefore, the resulting surface integral is defined on a space-like slice of the asymptotic AdS boundary $\partial V$ and the standard notation is
$$
 {\rm d} \Sigma_{\mu\nu} \propto \frac1{\det(g_{\mu\nu})} \epsilon_{\mu\nu\rho\sigma\gamma\delta} {\rm d} x^\rho \wedge {\rm d} x^\sigma \wedge {\rm d} x^\gamma \wedge {\rm d} x^\delta \, .
$$}
\be
 \label{eq:QQcommutator}
 \{ {\cal Q}, {\cal Q} \} \propto \int_{\partial V} {\rm d} \Sigma_{\mu\nu} \, \epsilon^{\mu\nu\rho\sigma\gamma\delta} \bar{\epsilon}^A \gamma_{\rho\sigma\gamma} \wt{D}_\delta \epsilon_A \, ,
\ee
where $\epsilon_A$ is the Killing spinor preserved by AdS$_6$, and the super-covariant derivative $\wt{D}$
includes all terms on the right hand side of the gravitino variation in \eqref{BPS}, \ie\;$\delta \psi_{A\mu} = \wt{D}_\mu \epsilon_A$.
The above anti-commutator is the explicit field dependent realization of the abstract AdS$_6$ superalgebra, ${\rm F}(4)$,
generating a combination of different asymptotic bosonic charges of the $S\O(5,2) \times \SU(2)_R$ generators. We are interested in the term in the gravitino variation in \eqref{BPS} proportional to $T_{(AB)}$, which precisely enters in the definition of the conserved $\SU(2)$ R-charge. We are further breaking the R-symmetry down to $\U(1)$ so we only need to look at the part proportional to $\sigma^3_{AB}$, \cf\;\eqref{T-tensor}. We are then led to the following formula for the conserved $\U(1)$ R-symmetry charge of a given solution,
\be
 \label{eq:R-symmetry}
 R_{\U(1)} \propto \int_{\partial V} {\rm d} \Sigma_{\mu\nu}\ e^{-\sigma} (L^{-1})_{r=3 | \Lambda} (F^{\Lambda})^{\mu\nu}\ .
\ee
Note that we are only interested to know the R-symmetry at a given radial slice of the spacetime
(that when interpreted as a holographic RG flow becomes a measure of how the R-symmetry changes along the flow),
not at the value of the asymptotic conserved charge.
Therefore, we can extract a normalized version of the integrand that we hope to match with the R-symmetry mixing in field theory.
Considering that $L^{-1}_{33} = \cosh (\phi_3)$, $L^{-1}_{3 4} = -\sinh (\phi_3)$ and that the democratic choice of $\U(1)$'s corresponds
to taking $F_{1, \mu\nu} \equiv F_{3, \mu\nu} + F_{4, \mu\nu}$, $F_{2, \mu\nu} \equiv F_{3, \mu\nu} - F_{4, \mu\nu}$ we finally define
\be
 \label{eq:R-symmetry-normalized}
 R_{\text{sugra}} \equiv X^1 F_1 + X^2 F_2 \, ,
\ee
where the mixing of the \emph{democratic} $\U(1)$ symmetries $F_{1,2}$ is given by the scalar dependent quantities
\bea
 \label{parametrization}
 \frac{X^1}{\pi} \equiv 1+ \tanh (\phi_3 ) \, ,
 \qquad \frac{X^2}{\pi} \equiv 1- \tanh (\phi_3 )\, , \qquad   e^{\phi_3} = \left( \frac{X^1}{X^2} \right)^{1/2} \, .
\eea

\section[The AdS\texorpdfstring{$_6$}{(6)} vacuum]{The AdS$_6$ vacuum}\label{AdS6}

The ${\rm F}(4)$ supergravity discussed in the previous section has an AdS$_6$ vacuum if we set $g=3m$ \cite{Romans:1985tw,Andrianopoli:2001rs,DAuria:2000xty}. 
Indeed, considering a background with metric
\be
 \rd s^2 = e^{2 f(r)} \left( \rd t^2 - \rd r^2 - \sum_{i=1}^4 \rd x_i^2 \right) \, ,
\ee
a nontrivial scalar profile for $\sigma(r)$ and $\phi_3(r)$, and setting all other fields to zero, the BPS equations \eqref{BPS} reduce to\footnote{One can derive these equations by taking the ultraviolet limit of the more general flow equations \eqref{BPS:AdSxSigma}, \eqref{BPS:AdSxSigmaxSigma}, or \eqref{BPS:AdSxM4}.}
\bea
 0 & = e^{-f} f' + \frac 12 \left( g \cosh (\phi_3) e^\sigma + m e^{-3 \sigma} \right) \, ,\\
 0 & = e^{-f} \sigma' - \frac 12 \left( g \cosh (\phi_3) e^\sigma - 3  m e^{-3 \sigma} \right) \, ,\\
 0 & =  e^{-f} \phi_3' -2 g \sinh (\phi_3) e^\sigma \, ,
\eea
where prime denotes the derivative of the function with respect to the radial coordinate $r$.
With $g=3m$, the AdS$_6$ background corresponds to $e^{-2f}=r^2$ and $\sigma = \phi_3=0$. 
We have further set $m=1/2$ so that the AdS$_6$ radius is normalized to one. 

A more suggestive way of solving the above equations is by taking the Ansatz $e^{2 f(r)}= e^{2 f_0} /r^2$, and $\sigma, \phi_3$ independent of $r$.
We can write the BPS equations in an alternative form by using the parameterization \eqref{parametrization}. The BPS equations for the fields in the gravity multiplet in terms of $X^{1,2}$ can be solved as 
\be
 e^\sigma = \frac{1}{\pi^{1/4}}  ( X^1 X^2 )^{1/8} \, , \qquad e^{f_0} = e^{3 \sigma} \, .
\ee
The on-shell supergravity action is given by
\be\label{prepotential}
 \cI_{\text{AdS}_6} (X^I) = -\frac{\pi^2 e^{4 f_0}}{3 G_{\text{N}}} = -  \frac{1}{3 \pi G_{\text{N}}} (X^1 X^2 )^{3/2} \, .
\ee
We then see that the BPS equation for $\phi_3$, which implies $\phi_3 = 0$, is equivalent to extremizing $\cI_{\text{AdS}_6}$ with respect to $X^I$.

The function $\cI_{\text{AdS}_6} (X^I)$ has a natural field theory interpretation.
The $S^5$ free energy of the $\USp(2N)$ theory reads  \cite{Jafferis:2012iv}
\be
 F_{S^5} = - \frac{9 \sqrt{2} \pi}{5} \frac{N^{5/2}}{\sqrt{8 - N_f}} \, .
\ee
This can be generalized to the case where a mass parameter is turned on for $\U(1) \subset \SU(2)_M$ \cite{Chang:2017mxc}%
\footnote{What we denote as chemical potentials here are actually mass parameters for the antisymmetric matter field in the $S^5$ free energy.
Comparing to \cite{Chang:2017mxc} we have $\Delta_1 = \pi \left( 1 + \frac{2 i}{3} m_{\text{as}} \right)$, $\Delta_2 = \pi \left( 1 - \frac{2 i}{3} m_{\text{as}} \right)$. }
\be
 F_{S^5} (\Delta_I) = - \frac{9 \sqrt{2}}{5 \pi^2} \frac{N^{5/2}}{\sqrt{8 - N_f}} \left( \Delta_1 \Delta_2 \right)^{3/2} \, ,
\ee
where $\Delta_1+\Delta_2= 2\pi$ and  the extremal value is recovered for $\Delta_1=\Delta_2=\pi$. Upon using the standard AdS$_6$/CFT$_5$ dictionary \cite{Jafferis:2012iv}
\be
 \label{AdS6-CF5:dict}
 G_{\text{N}} = \frac{5\pi}{27\sqrt{2}} \frac{\sqrt{8 - N_{f}}}{N^{5/2}} \, ,
\ee
and identifying $X^I \equiv \Delta_I$, we find that
\be
 F_{S^5} (\Delta_I) = \cI_{\text{AdS}_6} (X^I) \, .
\ee
Interestingly, as shown in \cite{Hosseini:2018uzp}, the same quantity is also related to the Seiberg-Witten prepotential of the five-dimensional theory on $\mathbb{R}^4 \times S^1$
which can be written as 
\be
 \cF (\Delta_I) = -\frac{2 \pi i} {27}F_{S^5} (\Delta_I) = - \frac{2 \pi i}{27} \cI_{\text{AdS}_6} (X^I)  \, .
\ee

As discussed in the introduction, the function $\cI_{\text{AdS}_6} (X^I)$ in six dimensions plays a role similar to the prepotential of four-dimensional $\cN = 2$ gauged supergravity.
In the AdS$_4$ black hole story, the supergravity prepotential is similarly related both to the twisted superpotential and to the $S^3$ free energy of the dual field theory \cite{Benini:2015eyy,Hosseini:2016tor,Hosseini:2017fjo}.%
\footnote{For black holes in AdS$_4\times S^7$ the prepotential is proportional to the function $\sqrt{X^1 X^2 X^3 X^4}$ with $\sum_{I=1}^4 X^I = 2\pi$ and for massive type IIA black holes to $(X^1 X^2 X^3)^{2/3}$ with $\sum_{I=1}^3 X^I = 2\pi$.}

\section[The AdS\texorpdfstring{$_4 \times \Sigma_\fg$}{(4) x Sigma(g)} solution]{The AdS$_4 \times \Sigma_\fg$ solution}\label{AdS4}

The ${\rm F}(4)$ gauged supergravity has also an AdS$_4 \times \Sigma_\fg$ solution corresponding to the twisted compactification of the five-dimensional SCFT
on a Riemann surface $\Sigma_\fg$ of genus $\fg$. In the infrared the field theory flows to a three-dimensional SCFT.

We consider the following Ansatz for the metric
\be
 \rd s^2 = e^{2 f(r)} \left( \rd t^2 -\rd r^2 -  \rd z_1^2 - \rd z_2^2 \right) - e^{2 h(r)} \rd s^2_{\Sigma_{\fg}} \, ,
\ee
and for the gauge fields $\U(1) \times \U(1) \subset \SU(2)_R \times \U(1)$:
\be
 F^{r=3} = \frac{\zeta}{g} \kappa\, \vol( \Sigma_\fg) \, ,\qquad F^{I=1} = \frac{\zeta}{g} p\,\vol ( \Sigma_\fg) \, ,
\ee
with $\zeta =\pm 1$. There is a nontrivial profile for the scalars $\sigma(r), \phi_3(r)$ and all other fields are set to zero.
Here, $\Sigma_{\fg}$ is a Riemann surface with metric normalized as $R_{\mu\nu} = \kappa g_{\mu\nu}$, with $\kappa=1$ for $S^2$, $\kappa=0$ for $T^2$, and $\kappa =-1$ for $\fg>1$.
With this normalization $\vol (\Sigma_\fg) = 2 \pi \eta_\fg$ with $\eta_\fg=2 |\fg -1|$ for $\fg \ne 1$ and $\eta_\fg=1$ for $\fg=1$.
The $\U(1) \subset \SU(2)_R$ gauge field is chosen in order to cancel the spin connection while the magnetic flux $p$ parameterizes a family of three-dimensional SCFTs.

If we choose spinors satisfying 
\be
 \label{twistprojection}
 \gamma^{34} \epsilon_A = - i \zeta\sigma^3_{AB} \epsilon^B \, ,
\ee
where the frame indices $3,4$ refer to the Riemann surface, the $\U(1) \subset \SU(2)_R$ gauge field cancels the spin connection along $\Sigma_{\fg}$.
This is precisely the topological twist. Requiring in addition that 
\be
 \gamma^{\hat r} \epsilon_A= - i \epsilon_A \, ,
\ee
where $\hat r$ is a frame index along the radial direction, the BPS equations \eqref{BPS} reduce to%
\footnote{Here we correct a numerical factor in the gaugino variation in \cite{Karndumri:2015eta}.}
\bea
 \label{BPS:AdSxSigma}
 0 & = e^{-f} f'  - \frac{1}{8 g} e^{-\sigma-2h} \left( \kappa \cosh (\phi_3) - p \sinh (\phi_3)\right) + \frac 12 \left(g \cosh (\phi_3) e^\sigma + m e^{-3 \sigma} \right) \, ,\\
 0 & = e^{-f} h'  + \frac{3 }{8 g} e^{-\sigma-2h} \left( \kappa \cosh (\phi_3) - p \sinh (\phi_3)\right) + \frac 12 \left(g \cosh (\phi_3) e^\sigma + m e^{-3 \sigma} \right) \, ,\\
 0 & = e^{-f} \sigma'  + \frac{1}{8 g} e^{-\sigma-2h} \left( \kappa \cosh (\phi_3) - p \sinh (\phi_3)\right)  - \frac 12 \left(g \cosh (\phi_3) e^\sigma - 3  m e^{-3 \sigma} \right) \, ,\\
 0 & =  e^{-f} \phi_3' + \frac{1}{2 g} e^{-\sigma-2h} \left( p \cosh (\phi_3) -  \kappa \sinh (\phi_3)\right)  - 2 g \sinh (\phi_3) e^\sigma \, .
\eea
We choose the parameterization of the scalar field $\phi_3$ as in \eqref{parametrization} and, in addition, we introduce a redundant but democratic parameterization for the flux 
\be\label{fluxesAdS4}
 \fs_1 \equiv 1- \fg + \frac{\eta_\fg}{2} p \, , \qquad \fs_2 \equiv 1 - \fg - \frac{\eta_\fg}{2} p \, ,
\ee
with $\fs_1 + \fs_2 = 2 (1 - \fg)$. We look for AdS$_4\times \Sigma_\fg$ vacua where $e^{f(r)}=e^{f_0}/r$ and $h(r), \sigma(r)$ and $\phi_3(r)$ are constant.  
Using the BPS equations \eqref{BPS:AdSxSigma}, the fields in the gravity multiplet can be solved in terms of the $X^I$'s as
\bea
e^\sigma = \left( \frac{2}{3 \pi} \right)^{1/4} ( X^1 X^2 )^{1/8} \, , \qquad e^{h} = \frac{1}{\sqrt{3 \eta_\fg}} \frac{ \left( - \fs_1 X^2 - \fs_2 X^1 \right)^{1/2}}{( X^1 X^2 )^{1/4}} e^{\sigma} \, , \qquad e^{f_0} =  e^{3 \sigma}  \, ,
\eea
where we set  $g=3m$ and $m=1/2$. 
The on-shell supergravity action can be written as 
\be
 \cI_{\text{AdS}_4} (X^I) = \frac{\pi e^{2 f_0 + 2 h} \vol (\Sigma_\fg)}{2 G_{\text{N}}} = - \frac{4}{27 G_{\text{N}}} (X^1 X^2 )^{1/2} \left( \fs_2 X^1 + \fs_1 X^2 \right) \, .
\ee
It turns out that the BPS equation for $\phi_3$ -- last line of \eqref{BPS:AdSxSigma} -- is equivalent to the extremization of $ \cI_{\text{AdS}_4} (X^I)$
with respect to $X^I$. 
The previous expression can be more elegantly rewritten as
\be\label{freeAdS4}
 \cI_{\text{AdS}_4} (X^I) = - \frac{8}{81 G_{\text{N}}} \sum_{I = 1}^2 \fs_I \frac{\partial \left( X^1 X^2 \right)^{3/2}}{\partial X^I} \, .
\ee
As expected, using \eqref{AdS6-CF5:dict} and identifying $X^I \equiv \Delta_I$, we find that
\be
 F_{S^3 \times \Sigma_\fg} (\Delta_I) = \cI_{\text{AdS}_4} (X^I) \, ,
\ee
where $ F_{S^3 \times \Sigma_\fg} $ is the $S^3 \times \Sigma_{\fg}$ free energy of the same theory, as a function of R-charges,  computed in \cite{Crichigno:2018adf}%
\footnote{Comparing to \cite{Crichigno:2018adf} we have $\fs_1 = (1 - \fg) (1 + \hat \fn_M)$, $\fs_2 = (1 - \fg) (1 - \hat \fn_M)$, $\Delta_1 = \pi ( 1 + \wt\nu_{\text{AS}})$, $\Delta_2 = \pi ( 1 - \wt\nu_{\text{AS}} )$.}
\be
 F_{S^3 \times \Sigma_\fg} (\Delta_I)  = - \frac{8  \sqrt{2}}{15 \pi} \frac{N^{5/2}}{\sqrt{8 - N_{f}}} \sum_{I = 1}^2 \fs_I \frac{\partial ( \Delta_1 \Delta_2)^{3/2}}{\partial \Delta_I} \, .
\ee
Here, $\Delta_1+\Delta_2=2\pi$. Moreover, as noticed in \cite{Hosseini:2018uzp}, this expression is also related to the effective twisted superpotential $\wt \cW$ of the theory compactified on $\Sigma_{\fg} \times S^1$: 
\be
 \wt \cW (\Delta_I)  = \frac{ \pi i}{2}  F_{S^3 \times \Sigma_\fg} (\Delta_I) =  \frac{ \pi i}{2}  \cI_{\text{AdS}_4} (X^I)  \, .
\ee

The extremization of $F_{S^3 \times \Sigma_\fg} (\Delta_I)$ with respect to $\Delta_I$  determines the exact R-symmetry of the three-dimensional field theory 
that is obtained by twisted compactification on $\Sigma_\fg$. The critical value of  $F_{S^3 \times \Sigma_\fg} (\Delta_I)$ is the free energy of the theory
and coincides with the value derived directly in ten-dimensional massive type IIA supergravity in \cite{Bah:2018lyv}. This is an evidence that
the gauged supergravity provides a consistent truncation of the ten-dimensional theory. 

\section[The AdS\texorpdfstring{$_2 \times \Sigma_{\fg_1} \times \Sigma_{\fg_2}$}{(4) x Sigma(g1) x Sigma(g2)} solution]{The AdS$_2 \times \Sigma_{\fg_1} \times \Sigma_{\fg_2}$ solution}\label{AdS2}

Now we search for black hole horizon solutions of the form AdS$_2 \times \Sigma_{\fg_1} \times \Sigma_{\fg_2}$. We consider the following Ansatz for the metric 
\be
 \rd s^2 = e^{2 f(r)} \left( \rd t^2 -\rd r^2  \right) - e^{2 h_1(r)} \rd s^2_{\Sigma_{\fg_1}} - e^{2 h_2(r)} \rd s^2_{\Sigma_{\fg_2}} \, ,
\ee
and the gauge fields 
\be
 F^{r=3} = \frac{\zeta}{g} \kappa_1\;\! \vol ( \Sigma_{\fg_1} ) + \frac{\zeta}{g} \kappa_2\;\! \vol ( \Sigma_{\fg_2} )  \, ,\qquad
 F^{I=1} = \frac{\zeta}{g} p_1\;\! \vol ( \Sigma_{\fg_1} ) + \frac{\zeta}{g} p_2\;\! \vol ( \Sigma_{\fg_2} ) \, ,
\ee
with $\zeta =\pm 1$ and the previous conventions for Riemann surfaces. The $\U(1) \subset \SU(2)_R$ gauge field is chosen in order to cancel the spin connection and $p_1$ and $p_2$ are magnetic charges, one for each Riemann surface.  
There is as usual a nontrivial profile for the scalars $\sigma(r), \phi_3(r)$. This time the two-form $B_{\mu \nu}$ cannot be set to zero. Assuming $H_{\mu\nu\lambda}=0$, the equations of motion require that 
\be
 e^{-2\sigma} m^2 {\cal  N}_{00}  B^{\mu\nu} +\frac{1}{16} \epsilon^{\mu\nu\tau\rho\lambda\sigma} \eta_{\Lambda\Sigma} F^\Lambda_{\tau\rho} F^\Sigma_{\lambda\sigma} = 0 \, ,
\ee
which is  solved by
\be
 B_{tr}= - \frac{ (p_1 p_2 - \kappa_1 \kappa_2)}{8 m^2 g^2} e^{2\sigma+2 f-2 h_1-2h_2} \, .
\ee
With the spinor projections 
\be
 \label{twistprojection2}
 \gamma^{12} \epsilon_A = - i \zeta\sigma^3_{AB} \epsilon^B \, , \qquad
 \gamma^{34} \epsilon_A = - i \zeta\sigma^3_{AB} \epsilon^B\, , \qquad
 \gamma^{\hat r} \epsilon_A= -i \epsilon_A \, ,
\ee
where the frame indices $1,2$ refer to the first Riemann surface and $3,4$ to the second, the $\U(1) \subset \SU(2)_R$ gauge field cancels the spin connection, and the  BPS equations \eqref{BPS} reduce to
 \begin{align}
\begin{split}
 \label{BPS:AdSxSigmaxSigma}
 0 & = e^{-f} f'  - \frac{1}{8 g} e^{-\sigma-2h_1} \left(\kappa_1 \cosh (\phi_3) - p_1 \sinh (\phi_3) \right) - \frac{1}{8 g} e^{-\sigma-2h_2} \left( \kappa_2\cosh (\phi_3) - p_2 \sinh (\phi_3)\right)  \\
 & +  \frac 12 \left(g \cosh (\phi_3) e^\sigma + m e^{-3 \sigma}\right)  
 - \frac{3 (p_1 p_2 - \kappa_1 \kappa_2)}{32 m g^2} e^{\sigma - 2 h_1-2 h_2} \, ,\\
 0 & = e^{-f} h_1'  + \frac{3 }{8 g} e^{-\sigma-2h_1} \left( \kappa_1\cosh (\phi_3) - p_1 \sinh (\phi_3)\right)  - \frac{1}{8 g} e^{-\sigma-2h_2} \left( \kappa_2\cosh (\phi_3) - p_2 \sinh (\phi_3)\right) \\
 & + \frac 12 \left(g \cosh (\phi_3) e^\sigma + m e^{-3 \sigma}\right)  
 +  \frac{ (p_1 p_2 - \kappa_1 \kappa_2)}{32 m g^2} e^{\sigma - 2 h_1-2 h_2} \, ,\\
 0 & = e^{-f} h_2'  - \frac{ 1}{8 g} e^{-\sigma-2h_1} \left( \kappa_1\cosh (\phi_3) - p_1 \sinh (\phi_3)\right)  + \frac{3}{8 g} e^{-\sigma-2h_2} \left(\kappa_2 \cosh (\phi_3) - p_2 \sinh (\phi_3)\right) \\
 & + \frac 12 \left(g \cosh (\phi_3) e^\sigma + m e^{-3 \sigma}\right)  
 + \frac{ (p_1 p_2 -\kappa_1 \kappa_2 )}{32 m g^2} e^{\sigma - 2 h_1-2 h_2} \, ,\\
 0 & = e^{-f} \sigma'  + \frac{1}{8 g} e^{-\sigma-2h_1} \left(\kappa_1 \cosh (\phi_3) - p_1 \sinh (\phi_3)\right) + \frac{1}{8 g} e^{-\sigma-2h_2} \left( \kappa_2\cosh (\phi_3) - p_2 \sinh (\phi_3)\right) \\
 & - \frac 12 \left(g \cosh (\phi_3) e^\sigma - 3  m e^{-3 \sigma}\right) 
 - \frac{ (p_1 p_2 - \kappa_1 \kappa_2)}{32 m g^2} e^{\sigma - 2 h_1 -2 h_2} \, ,\\
 0 & = e^{-f} \phi_3' + \frac{1}{2 g} e^{-\sigma-2h_1} \left( p_1 \cosh (\phi_3) -  \kappa_1\sinh (\phi_3)\right) + \frac{1}{2 g} e^{-\sigma-2h_2} \left( p_2 \cosh (\phi_3) -  \kappa_2 \sinh (\phi_3)\right)  \\
 & - 2 g \sinh (\phi_3) e^\sigma \, .
\end{split}
\end{align}
We choose the parameterization \eqref{parametrization} for the scalar field $\phi_3$ and a democratic parameterization for the fluxes 
\bea\label{parfluxes}
 \fs_1 \equiv 1 - \fg_1 + \frac{\eta_{\fg_1}}{2} p_1 \, , \qquad  \fs_2 \equiv 1 - \fg_1 - \frac{\eta_{\fg_1}}{2} p_1 \, , \\
 \ft_1 \equiv 1 - \fg_2 + \frac{\eta_{\fg_2}}{2} p_2 \, , \qquad  \ft_2 \equiv 1 - \fg_2 - \frac{\eta_{\fg_2}}{2} \, p_2 \, , 
\eea
with  $\fs_1 + \fs_2 = 2 ( 1 - \fg_1)$ and $\ft_1 + \ft_2 = 2 ( 1- \fg_2)$.
To have a black hole horizon AdS$_2 \times \Sigma_{\fg_1}\times \Sigma_{\fg_2}$ we set $e^{f(r)}=e^{f_0}/r$ and $h_1(r), h_2(r), \sigma(r)$ and $\phi_3(r)$ constant.  
Using the BPS equations \eqref{BPS:AdSxSigmaxSigma}, the fields in the gravity multiplet can be solved in terms of $X^I$ as
\bea
 & e^\sigma = (X^1 X^2)^{1/8} \left( \frac{ ( \fs_1 X^2 + \fs_2 X^1)(\ft_1 X^2 + \ft_2 X^1) + 2 X^1 X^2 (\fs_2 \ft_1 + \fs_1 \ft_2)}{3 \pi( \fs_1 X^2 + \fs_2 X^1)(\ft_1 X^2 + \ft_2 X^1)} \right)^{1/4} , \\
 & e^{f_0} = \frac{( X^1 X^2)^{1/2}}{3 \pi} e^{- \sigma} \, , \\
 & e^{h_1} = \frac{1}{\sqrt{3 \eta_{\fg_1}}} \frac{ \left( -\fs_1 X^2 - \fs_2 X^1 \right)^{1/2}}{( X^1 X^2 )^{1/4}} e^{\sigma} \, , \\
 & e^{h_2} = \frac{1}{\sqrt{3 \eta_{\fg_2}}} \frac{ \left( -\ft_1 X^2 - \ft_2 X^1 \right)^{1/2}}{( X^1 X^2 )^{1/4}} e^{\sigma}\, , 
\eea
where we set $g=3m$ and $m=1/2$. One can find families of regular horizons, with fluxes satisfying all the quantization conditions, whenever $\kappa_1=-1$ or $\kappa_2=-1$.  The Bekenstein-Hawking entropy can be written as
\be
 \label{entropyAdS2}
 \cI_{\text{AdS}_2} (X^I) = \frac{e^{2 h_1+2 h_2} \vol (\Sigma_{\fg_1}\times \Sigma_{\fg_2})}{4 G_{\text{N}}}
 = \frac{4 \pi}{81 G_{\text{N}}} \sum_{I , J =1}^2 \fs_I \ft_J \frac{\partial^2 \left( X^1 X^2 \right)^{3/2}}{\partial X^I \partial X^J} \, ,
\ee
as a function of $X^I$. It is quite remarkable that the BPS equation for $\phi_3$ -- last line of \eqref{BPS:AdSxSigmaxSigma} -- is equivalent to the extremization of $\cI_{\text{AdS}_2}(X^I)$
with respect to $X^I$. This is the \emph{attractor mechanism} in six-dimensional gauged supergravity:
once the fields in the gravity multiplet are expressed in terms of the scalars in the vector multiplet,
the entropy is obtained by extremizing the functional $\cI_{\text{AdS}_2}(X^I)$.

We can now compare the entropy of the six-dimensional black holes with the prediction of the topologically twisted index computed in \cite{Hosseini:2018uzp}.
The index, at large $N$, is given by \cite{Hosseini:2018uzp}
\be
 \cI_{\text{SCFT}} (\Delta_I) = \frac{4 \sqrt{2}}{15} \frac{N^{5/2}}{\sqrt{8 - N_{f}}} \sum_{I , J = 1}^2 \fs_I \ft_J \frac{\partial^2 ( \Delta_1 \Delta_2)^{3/2}}{\partial \Delta_I \partial \Delta_J} \, .
\ee
The index depends on a chemical potential $\Delta$ for the $\U(1)$ subgroup of the $\SU(2)$ global symmetry.
As in \cite{Hosseini:2018uzp}, we find it convenient to use a pair of redundant but democratic parameters
\be
 \label{dem}
 \Delta_1 = \Delta\, ,\qquad \Delta_2 = 2 \pi - \Delta \, ,
\ee
with $\Delta_1+\Delta_2= 2 \pi$.
In the spirit of the microscopic counting for magnetically charged AdS black holes in four dimensions, we expect that the entropy is obtained by extremizing $\cI_{\text{SCFT}} (\Delta_I)$ with respect to $\Delta_I$.
This was called $\cI$-extremization principle in \cite{Benini:2015eyy,Benini:2016rke}.  Using \eqref{AdS6-CF5:dict} and identifying $X^I \equiv \Delta_I$, we find that
\be
 \cI_{\text{SCFT}} (\Delta_I) = \cI_{\text{AdS}_2} (X^I) \, ,
\ee
and we see that the field theory $\cI$-extremization precisely corresponds to the attractor mechanism in supergravity.

\section[The AdS\texorpdfstring{$_2 \times \cM_4$}{(2) x M(4)} solution]{The AdS$_2 \times \cM_4$ solution}\label{AdS2M}

It is easy to find more general black hole horizons with Abelian twists. We consider the following metric
\be
 \rd s^2 =  e^{2 f(r)} \left(  \rd t^2 - \rd r^2\right) - e^{2 h(r)} \rd s^2_{\cM_4} \, ,
\ee
where $\cM_4$ is a K\"ahler-Einstein manifold with metric normalized as $R_{\mu\nu}=\kappa g_{\mu\nu}$ $(\kappa=\pm 1, 0)$, and gauge fields 
\be
 F^{r=3} = \frac{\zeta}{g} \kappa\, ( e^{12}+e^{34}) e^{-2 h(r)}  \, ,\qquad
 F^{I=1} = \frac{\zeta}{g} p\, ( e^{12}+e^{34}) e^{-2 h(r)} \, ,
\ee
where $e^i$, $i=1,2,3,4$, are vierbeins in the directions corresponding to the manifold $\cM_4$. The reduced holonomy group on the manifold, $\U(2)$, splits into $\U(1)$ that we choose to correspond to the selfdual part of the spin connection, $\omega^+$, and $\SU(2)$ for the anti-selfdual part, $\omega^-$.  
As in the previous section, there is  a nontrivial profile for the scalars $\sigma(r), \phi_3(r)$ and the two-form
\be
 B_{tr} = - \frac{ (p^2 - \kappa^2)}{8 m^2 g^2} e^{2\sigma+2 f-2 h_1-2h_2} \, .
\ee
With the spinor projections 
\be
 \label{twistprojection2}
 \gamma^{12} \epsilon_A = - i \zeta\sigma^3_{AB} \epsilon^B \, , \qquad  \gamma^{34} \epsilon_A = - i \zeta\sigma^3_{AB} \epsilon^B \, , \qquad \gamma^{\hat r} \epsilon_A= -i \epsilon_A \, ,
\ee
the $\U(1) \subset \SU(2)_R$ gauge field cancels the $\omega^+$ spin connection, while the $\omega^-$ part drops out of the Killing spinor covariant derivative, since \eqref{twistprojection2} imply $\gamma^{1234}\epsilon_A= -\epsilon_A$. The BPS equations \eqref{BPS} reduce to
\bea
 \label{BPS:AdSxM4}
 0 & = e^{-f} f'  - \frac{1}{4 g} e^{-\sigma-2h} \left(\kappa \cosh (\phi_3) - p \sinh (\phi_3)\right) 
 + \frac 12 \left(g \cosh (\phi_3) e^\sigma + m e^{-3 \sigma}\right) - \frac{3 (p^2 - \kappa^2)}{32 m g^2} e^{\sigma - 4 h} \, ,\\
 0 & = e^{-f} h'  + \frac{1 }{4 g} e^{-\sigma-2h} \left( \kappa\cosh (\phi_3) - p \sinh (\phi_3)\right)  + \frac 12 \left(g \cosh (\phi_3) e^\sigma + m e^{-3 \sigma}\right) + \frac{ (p^2 - \kappa^2)}{32 m g^2} e^{\sigma - 4 h} \, ,\\
 0 &= e^{-f} \sigma'  + \frac{1}{4 g} e^{-\sigma-2h} \left(\kappa \cosh (\phi_3) - p \sinh (\phi_3)\right)  - \frac 12 \left(g \cosh (\phi_3) e^\sigma - 3  m e^{-3 \sigma}\right) -  \frac{ (p^2 - \kappa^2)}{32 m g^2} e^{\sigma - 4 h} \, ,\\
 0 &=  e^{-f} \phi_3' + \frac{1}{ g} e^{-\sigma-2h} \left( p \cosh (\phi_3) -  \kappa\sinh (\phi_3)\right)   - 2 g \sinh (\phi_3) e^\sigma \, .
\eea 
We choose the parameterization \eqref{parametrization} for the scalar field $\phi_3$ and a democratic parameterization for  the fluxes 
\bea\label{fluxesM4}
 \fs_1 \equiv \kappa  + p  \, , \qquad  \fs_2 \equiv \kappa - p \, ,
\eea
with $\fs_1 + \fs_2 = 2 \kappa$. To have an AdS$_2 \times \cM_4$ horizon topology we set $e^{f(r)}=e^{f_0}/r$ and $h(r), \sigma(r)$ and $\phi_3(r)$ constant.  
Using the BPS equations \eqref{BPS:AdSxM4}, the fields in the gravity multiplet can be solved in terms of $X^I$ as
\bea
 & e^\sigma = (X_1 X_2)^{1/8} \left( \frac{ ( \fs_1 X^2 + \fs_2 X^1)^2 + 4 X^1 X^2 \fs_1 \fs_2 }{3 \pi( \fs_1 X^2 + \fs_2 X^1)^2} \right)^{1/4} \, , \\
 & e^{h} = \frac{1}{\sqrt{6}} \frac{ \left( -\fs_1 X^2 - \fs_2 X^1 \right)^{1/2}}{( X^1 X^2 )^{1/4}} e^{\sigma} \, ,
 \qquad e^{f_0} = \frac{( X^1 X^2)^{1/2}}{3 \pi} e^{- \sigma} \, ,
\eea
where we set  $g=3m$ and $m=1/2$. One can find  regular horizons  only for $\kappa = -1$ for sufficiently small $p$.\footnote{Since fluxes must be quantized, this condition puts some restriction on the choice of $\cM_4$.} 
The Bekenstein-Hawking entropy is then obtained by extremizing
\be\label{entropyM}
 \cI_{\text{AdS}_2} (X^I) = \frac{e^{4 h} \vol (\cM_4)}{4 G_{\text{N}}} = \frac{\vol (\cM_4)}{ 324 \pi G_{\text{N}}} \sum_{I , J =1}^2 \fs_I \fs_J \frac{\partial^2 \left( X^1 X^2 \right)^{3/2}}{\partial X^I \partial X^J} \, ,
\ee
with respect to $X^I$. Remarkably, the extremization of \eqref{entropyM} is equivalent to the BPS equation for $\phi_3$ -- last line of \eqref{BPS:AdSxM4}.
This formula is very simple and suggests that  the computation in \cite{Hosseini:2018uzp}
could be generalized to this case too. It is also reminiscent of similar expressions for AdS$_4$ black holes (see \cite{Hosseini:2018uzp} for a discussion). 

\section*{Acknowledgements}

We would like to thank Minwoo Suh for very useful discussions.
We also like to acknowledge the collaboration with Itamar Yaakov on related topics.
The work of SMH was supported by World Premier International Research Center Initiative (WPI Initiative), MEXT, Japan.
KH is supported in part by the Bulgarian NSF grant DN08/3 and the bilateral grant STC/Bulgaria-France 01/6.
The work of AP is supported by the Knut and Alice Wallenberg Foundation under grant Dnr KAW 2015.0083.
AZ is partially supported by the INFN and ERC-STG grant 637844-HBQFTNCER.
SMH would like to thank the Yau Mathematical Sciences Center of Tsinghua University in Beijing for their kind hospitality during his visit, where part of this work was done.

\begin{appendix}

\section{Conventions}
\label{sec:app}

We collect here few relevant conventions and formulae used in \cite{Andrianopoli:2001rs}. We refer to that paper for everything missing or forgotten.
Indices $A,B,\ldots $ in the fundamental representation of $\SU(2)_R$ are raised and lowered as $T^A = \epsilon^{AB} T_B$ and $ T_A= T^B \epsilon_{BA}$.
Indices $\Lambda,\Sigma,\ldots$ of $\SO(4,n_\text{V})$ are raised and lowered with $\eta_{\Lambda\Sigma} ={\rm diag}\{ 1,1,1,1,-1,\ldots, -1\}$. Spinors are pseudo-Majorana
with $(\psi_A)^\dagger \gamma^0 = \epsilon^{AB} \psi_B^t$ and $\gamma^7 = i \gamma^0\gamma^1 \gamma^2\gamma^3\gamma^4\gamma^5\gamma^6$. The potential $V$ and the fermionic shifts are constructed using the quantities
\be
 A = \epsilon^{rst} K_{rst} \, , \qquad B^i = \epsilon^{ijk} K_{jk0} \, , \qquad C^t_I=\epsilon^{trs} K_{rIs} \, , \qquad D_{It} =K_{0It} \, ,
\ee
where
\bea
& K_{rs\alpha} = g \epsilon_{lmn} L^{l}_{\phantom{I} r} (L^{-1})_s^{\phantom{s} m} L^n_{\phantom{n} \alpha} + g^\prime C_{IJK} L^{I}_{\phantom{I} r} (L^{-1})_s^{\phantom{s} J} L^K_{\phantom{K} \alpha} \, ,\\
& K_{\alpha It} = g \epsilon_{lmn} L^{l}_{\phantom{I} \alpha} (L^{-1})_I^{\phantom{I} m} L^n_{\phantom{n} t} + g^\prime C_{LJK} L^{L}_{\phantom{L} \alpha} (L^{-1})_I^{\phantom{I} J} L^K_{\phantom{K} t} \, .
\eea
Here, $C_{IJK}$ are the structure constants of the gauge group $G \subset \SO(n_\text{V})$.
For us $C_{IJK}=0$. We then have the following fermionic shifts entering the BPS equations \eqref{BPS}:
\bea
 S_{AB} & = \frac {i}{24} \left( A e^\sigma + 6 m e^{-3\sigma} (L^{-1})_{00} \right) \epsilon_{AB} -\frac i8 ( B_t e^\sigma -2  m e^{-3\sigma} (L^{-1})_{i0}) \gamma^7 \sigma^t_{AB} \, , \\
 N_{AB} & = \frac {1}{24} \left( A e^\sigma -18 m e^{-3\sigma} (L^{-1})_{00} \right)\epsilon_{AB} +\frac 18 ( B_t e^\sigma +6   m e^{-3\sigma} (L^{-1})_{i0}) \gamma^7 \sigma^t_{AB} \, ,\\
 M_{AB}^I & = \left( - C_t^I + 2 i \gamma^7 D_t^I \right) e^\sigma \sigma^t_{AB} - 2 m e^{-3 \sigma}  (L^{-1})^I_{\phantom{I} 0} \gamma^7 \epsilon_{AB} \, .
\eea
Finally, the scalar potential reads
\be
 V = -e^{2\sigma} \left( \frac {A^2}{36}  + \frac {B^i B_i}{4} + \frac {C^I_{\phantom{I} t} C_{It}}{4}  + D^I_{\phantom{I} t} D_{It} \right) + m^2 e^{-6 \sigma} {\cal N}_{00} - m e^{-2\sigma} \left( \frac 23 A L_{00} -2 B^i L_{i0} \right) .
\ee
\end{appendix}
\bibliographystyle{ytphys}

\bibliography{6D_mBHs}

\end{document}